\documentclass[12pt,preprint]{aastex}
\usepackage{natbib}
\usepackage{amsmath}
\def\ovi{{{\rm O}\,{\sc vi}~}}
\def\ovii{{{\rm O}\,{\sc vii}~}}
\def\oviii{{{\rm O}\,{\sc viii}~}}
\def\oviin{{{\rm O}\,{\sc vii}}}

\def\civ{{{\rm C}\,{\sc iv}~}}
\def\siiv{{{\rm Si}\,{\sc iv}~}}
\def\chandra{{\it Chandra}~}
\def\xmm{{\it XMM-Newton}~}
\def\suzaku{{\it Suzaku}~}

\def\b{{\it b}}

\def\msun{{M_{\odot}}}
\def\gax{${_>\atop^{\sim}}$}

\begin{document}

\title{A huge reservoir of ionized gas around the Milky Way: 
Accounting for the Missing Mass?}
\author{A. Gupta and S. Mathur\altaffilmark{1}}
\affil{Astronomy Department, Ohio State University, Columbus, OH 43210, USA}
\email{agupta@astronomy.ohio-state.edu}
\author{Y. Krongold}
\affil{Instituto de Astronomia, Universidad Nacional Autonoma de Mexico, Mexico City, (Mexico)}
\author{F. Nicastro}
\affil{Harvard-Smithsonian Center for Astrophysics, Cambridge, MA, 02138, USA}
\affil{Osservatorio Astronomico di Roma-INAF, 
Via di Frascati 33, 00040, Monte Porzio Catone, RM, (Italy)}
\author{M. Galeazzi}
\affil{Physics Department, University of Miami, Coral Gables, FL 33124, USA}

\altaffiltext{1}{Center for Cosmology and
Astro-Particle Physics, The Ohio State University, Columbus, OH 43210}

\begin{abstract}

Most of the baryons from galaxies have been ``missing'' and several
studies have attempted to map the circumgalactic medium (CGM) of
galaxies in their quest. We report on X-ray observations made with the
Chandra X-ray Observatory probing the warm-hot phase of the CGM of our
Milky Way at about $10^6~$K. We detect \ovii and \oviii absorption lines
at $z=0$ in extragalactic sight lines and measure accurate column
densities using both K$\alpha$ and K$\beta$ lines of \oviin. We then
combine these measurements with the emission measure of the Galactic halo
from literature to derive the density and the pathlength of the CGM.  We
show that the warm-hot phase of the CGM is massive, extending over a
large region around the Milky Way, with a radius of over $100~$kpc. The
mass content of this phase is over ten billion solar masses, many times
more than that in cooler gas phases and comparable to the total baryonic
mass in the disk of the Galaxy. The missing mass of the Galaxy appears
to be in this warm-hot gas phase.

\end{abstract}

\section{Introduction}

We have known for a while that the baryonic mass of galaxies, including
that of our own Milky way, falls short of what is expected for their
total mass (\cite{SL06,Bregman07} and references therein). 
This ``missing'' mass
is either ejected from galaxies into the intergalactic medium (IGM) or
still resides in the circumgalactic medium (CGM). Our galaxy is a member
of the Local Group of galaxies, and it is possible that the matter
ejected from the Galaxy resides in the Local Group. Either in the CGM or
in the Local Group medium (LGM), the baryonic mass is expected in a
warm-hot gas phase at temperatures between $10^{5}-10^{7}~K$. This is
probably the reason why earlier studies probing only cooler gas could
not account for it. The warm component of this phase at
$10^{5}-10^{6}~K$ has been observed in the UV traced by absorption lines
of ionized metals, in particular \ovi 
\cite{Nicastro2003,Sembach2003}. Recent observations with the Hubble
Space Telescope (HST) have shown that the CGM around star-forming
galaxies is large ($150$ kpc) and the mass of the warm phase traced
by \ovi exceeds that of the gas in the galaxies themselves
\cite{Tumlinson11,Tripp11}. Using other absorption lines, e.g. 
\siiv and \civ in the UV \cite{LH11} found that there is a large reservoir of 
warm ionized gas around our Galaxy as well.

The hotter phase of the warm-hot gas, at temperatures $10^{6}-10^{7}~K$
can be probed by even more highly ionized metals. The dominant
transitions from such ions lie in the soft-X-ray band and indeed,
absorption lines due to \ovii and \oviii at redshift zero have been
detected toward extragalactic sight-lines by the Chandra X-ray
Observatory and the XMM-Newton Observatory \cite{Nicastro2005,
Williams2005, Williams2007, Bregman2007}. The distribution, spacial 
extent, and mass
of the warm-hot gas provide important constraints to the models of large
scale structure formations (e.g., \cite{Cen1999}) and/or
outflows/inflows from galaxies (e.g. \cite{Stinson11}), but these
parameters are difficult to measure in part due to the limited spectral
resolution of \chandra and \xmm and in part because of the inherent
difficulty in using the absorption line studies alone. Here we use both
absorption and emission observations to derive the physical properties
of the warm-hot plasma.  We show that the X-ray observations probe
million degree gas, with low density, extending over $100$ kpc and
having mass over ten billion solar masses. This is several times more
than previously found in the CGM of the Milky Way. Alternatively, the
warm-hot gas we probe is from the extended LG medium.

\section{Sample selection \& Data reduction}

We search for and measure the absorption lines from highly ionized gas
at $z\sim0$, as seen in the spectra of most of the extragalactic
sight-lines toward bright active galactic nuclei (AGNs). We use high
signal-to-noise ratio observations with \chandra High Energy
Transmission Grating (HETG) or Low Energy Transmission Grating (LETG)
with focus on \ovii and \oviii absorption lines at $\lambda=21.602~$\AA\
and $\lambda=18.967~$\AA\ respectively. 

We began by selecting all the Chandra grating observations of AGNs that
were publicly available as of 2011 October 30 and that had exposure time
of at least $100~ks$.  This resulted in multiple observations of $50$
targets with both grating spectrometers aboard Chandra (HETG 
and LETG). 

We reduced the data using the standard Chandra Interactive Analysis of
Observations (CIAO) software (v4.3) and Chandra Calibration Database
(CALDB, v4.4.2) and followed the standard Chandra data reduction
threads. For the Chandra ACIS/HETG and ACIS/LETG observations, 
we co-added the
negative and positive first-order spectra and built the effective area
files (ARFs) for each observations using the \emph{fullgarf} CIAO
script.  Those pertaining to the ACIS/LETG observations were corrected
for the ACIS quantum efficiency degradation. Unlike ACIS, the HRC does
not have the energy resolution to sort individual orders, and each
spectrum contains contributions from all the diffraction orders.  For
the HRC/LETG observations we used the standard ARF files for orders 1
through 6 and convolved it with the relevant standard redistribution
matrix file (RMF). For the targets with multiple observations, we added
the grating spectra and averaged the associated ARFs using the ciao
script \emph{add\_grating\_spectra}, to increase the signal to noise
ratio of the spectra.  We only add the observations with the same
instrumental configuration and the observations made with different
instruments are analyzed separately.

Of the 50 sight lines initially selected, 29 have good enough
signal-to-noise ratios (S/N) near $21.602~\AA$ to detect \ovii absorption
lines.  The strong \ovii K$\alpha$ absorption lines near $21.602~\AA$
are clearly visible in 21 out of 29 sources with good signal-to-noise
spectra; thus the covering fraction of the \ovii systems is
$21/29=0.72$. In ~30\% of the sources significant \oviii K$\alpha$ lines
near $18.967~\AA$ are also observed.  Here we consider a sub-sample of
$8$ targets (Table 1), where both \ovii and \oviii K$\alpha$ local
absorption has been confidently detected (we have not included 3C273 in
this sample of 29 sources because the $z=0$ absorption 
lines in this sight-line might
be from a nearby supernova remnant). The detailed analysis of the
complete sample will be presented in a forthcoming paper (Gupta et
al. 2012, in preparation).  The local ($z\sim0$) \ovii and \oviii
absorption lines in $3$ of our $8$ targets have been reported previously
by other authors, but to ensure the consistency of data analysis, we
reanalyzed all the data and obtained the fit results independently; for
the other five targets we present new detections of $z\sim0$ lines.

\section{Analysis}

\subsection{Continuum, Intrinsic Absorption, and Emission}

The HETG and LETG spectra are binned to $0.01~\AA$ and $0.025~\AA$
respectively and analyzed using the CIAO fitting package
\emph{Sherpa}. Since we are interested in oxygen absorption lines we fit
the spectral continuum in the $17-23~\AA$ range with a simple power law
and a Galactic absorbing column density (N$_{H}$, from \cite{DL1990}).
For the targets Mrk290 and NGC3783 we need to add a black-body component
to account for the soft excess, that could come from a standard
optically-thick accretion disk. In our sample, the photon index of the
power-law varies from $1.4-2.1$ with average value of $2.0$.


Our targets are nearby Type 1 AGNs, which have their own intrinsic
absorption and emission features. We carefully study all the spectra to
confirm that none of the AGN intrinsic features contaminates the
local ($z=0$) absorption lines. Except for one source, NGC4051, the
intrinsic lines are sufficiently red-shifted that they do not contaminate
the local \ovii and \oviii absorption lines. For this reason we do not
include NGC4051 in our final sample. We then modeled all the
statistically significant AGN intrinsic absorption and emission features
with Gaussian components.

\subsection{Local ($z\sim0$) Absorption}

After fitting the continuum and intrinsic features as described above,
the local \ovii and \oviii K$\alpha$ absorption lines are detected with
$\geq~3\sigma$ and $\geq~2\sigma$ significance levels respectively.  In
6 sources, we also detected the \ovii $K\beta$ absorption line near
$18.62\AA$. We fit these lines in {\it Sherpa} with narrow Gaussian
features.  Since with Chandra gratings ($FWHM=0.05~\AA$ for LEG and
$FWHM=0.023~\AA$ for MEG) the lines are unresolved, we fix the line
width to $1~m\AA$.  Errors are calculated using the \emph{projection}
command in {\it Sherpa}, allowing the overall continuum normalization to
vary along with all parameters for each line.  For the observations with
no detection of \ovii $K\beta$, we fixed the line centroid at
$18.629~\AA$ and obtained the upper limits on equivalent widths
(EW). The best-fit line equivalent
widths (EWs) and statistical uncertainties are given in Table 2 and the
spectra are shown in figure 1.

\subsection{Column Density Measurement}

For optically thin gas the ionic column density depends simply on the
observed equivalent width: $N(ion)=1.3\times
10^{20}(\frac{EW}{f\lambda^2})$, where $N(ion)$ is the ionic column
density ($cm^{-2}$), $EW$ is the equivalent width ($\AA$), $f$ is the
oscillator strength of the transition, and $\lambda$ is in $\AA$.
However, at the measured column densities of N(\oviin), saturation could
be an important issue as suggested by simulations \cite{Chen2003} and
observational studies of Mrk421 \cite{Williams2005}. Therefore to
correctly convert the measured equivalent widths to ionic column
densities, we require knowledge of the Doppler parameter \emph{b}; at a
fixed EW, column density decreases with increasing \emph{b}. The low
velocity resolution of \chandra gratings makes it unfeasible to directly
measure the \ovii line width. If multiple absorption lines from the same
ion are detected, the relative equivalent widths of these lines can
instead be used to place limits on the column density N(\oviin) and the
Doppler parameter \emph{b} of the medium.

To use this technique, we searched for and detected \ovii $K\beta$ line
at $18.629~$\AA\ in 6 out of 8 targets and measured the upper limit for
the remaining two.  For \oviin, the expected
$\frac{EW(K\beta)}{EW(K\alpha)}$ ratio is $\frac{f(K\beta)\times
\lambda^2(K\beta)}{f(K\alpha)\times \lambda^2(K\alpha)}=0.156$. Our
observations indicate that most \ovii K$\alpha$ lines are saturated
(Table 2). To place quantitative constraints on N(\ovii) and the
\emph{b} parameter we employ the technique described in details in
\cite{Williams2005}. For a given absorption line with a measured
equivalent width and known oscillator strength \emph{f} value, the
inferred column density as a function of the \emph{b} parameter can be
calculated using the relations from \cite{Spitzer1978}.  The \emph{b}
and N(\ovii) can be determined for a range of Doppler parameters for
which both K$\alpha$ and K$\beta$ transitions provide consistent
N(\ovii) measurements. Figure 2 shows such $1\sigma$ contours for the
measured \ovii K$\alpha$ and K$\beta$ transitions for NGC3783. As the
figure show, the $1\sigma$ constraints on Doppler parameter \emph{b} and
\ovii column densities are $45<b<128~km~s^{-1}$ and
$16.03<$logN(\ovii)$<16.68~cm^{-2}$, respectively.  For Mrk421 and
PKS2155-304 we used the \emph{b} parameter values from
\cite{Williams2005,Williams2007}.  Following the same method, we
constrained the \emph{b} parameter and \ovii column densities for all
other observations. Our measured column density toward Mrk 421 is
consistent with that in \cite{Yao2008}.  The upper limit on the
column density through the Galactic halo by Yao et al. is based on
several assumption which, in our opinion, are faulty, but a detailed
discussion is beyond the scope of this {\it Letter}. The \ovii column
densities for our sample range from $\log N$(\ovii)$=15.82$ to $16.50$
cm$^{-2}$, with a weighted mean value of $\log N$(\oviin)$=16.19\pm0.08$
cm$^{-2}$.  These values of N(\ovii) are higher than measured by other
authors \cite{Fang2006, Bregman2007} who assumed the lines to be
unsaturated.

\section{Results}

Column density ratios of different ions of the same element 
depend on the physical state of the medium and provide 
the rigorous constraints on the temperature of the absorbing medium. 
Given that we detect both \ovii and 
\oviii, we place a tight constraint on the temperature 
$\log T=6.1-6.4~K$, assuming the gas to be in 
collisional ionization equilibrium.

With the spectral resolution of current X-ray gratings it is difficult 
to resolve the absorption lines into Galactic and Local Group 
components. However a comparison between emission and absorption measurements, 
provides us with a great tool to analyze the 
properties of the X-ray absorbers. The absorption lines 
measure the column density of gas $N_H= \mu n_e R$, 
where $\mu$ is the mean molecular weight $\approx 0.8$, $n_e$ is 
the electron density and $R$ is the path-length.  The emission measure, 
on the other hand, is sensitive to the square of the number density of 
the gas ($EM= n_e^2 R$, assuming a constant density plasma).  Therefore 
a combination of absorption and emission measurements naturally 
provides constraints on the density and the path-length of the 
absorbing/emitting plasma.

\cite{Henley2010} and \cite{Yoshino2009} using \xmm and \suzaku data
respectively, found that the Galactic halo temperature is fairly
constant across the sky $(1.8-2.4)\times10^{6}~K$, but the halo emission
measure varies by an order of magnitude $(0.0005-0.005~cm^{-6}~pc)$ with
an average of $EM=0.0030\pm0.0006~cm^{-6}~pc$, assuming solar
metallicity\footnote{The X-ray spectrum of the diffuse Galactic halo
emission is normally fitted with collisional ionization equilibrium
(CIE) plasma models (e.g. APEC in XSPEC). The free parameters of this
model are temperature, metallicity and the emission measure. In the
low-resolution spectrum the emission measure and metallicity cannot be
fit independently, so the EM is obtained for a given metallicity;
smaller the metallicity, larger the EM.}.  Other measurements of
Galactic Halo emission \cite{McCammon2002, Galeazzi2007, 
Gupta2009,Hagihara2010} also reported
$EM\sim 0.003~cm^{-6}~pc$ for solar metallicity, close to the average. 
Thus using $EM=0.003~(\frac{Z_{\odot}}{Z})(\frac{8.51 \times 10^{-4}}{(A_O/A_H)})~cm^{-6}~pc$, we solve for the
path length and electron density of the absorbing gas. Combining
absorption and emission the density is:

\begin{equation}
n_e= (2.0\pm0.6 \times 10^{-4}) (\frac{0.5}{f_{O VII}})^{-1} cm^{-3}
\end{equation}

and the path length: 

\begin{equation}
R = (71.8\pm30.2) (\frac{8.51 \times 10^{-4}}{(A_O/A_H)}) 
(\frac{0.5}{f_{O VII}})^2 (\frac{Z_{\odot}}{Z}) ~kpc
\end{equation}

where the Solar Oxygen abundance of $A_O/A_H= 8.51 \times 10^{-4}$ is
from \cite{AG89}, $f_{O VII}$ is the ionization fraction of \ovii and
$Z$ is the metallicity. Newer values of oxygen
abundance are even lower \cite{Holweger01,Asplund2009}; making $L$
larger. For the observed temperature of about
\gax$10^6~$K, it is reasonable to expect $f=0.5$ (see, e.g., figure 4 in
\cite{Mathur03}). Cosmological simulations \cite{Toft2002,SL06} of
formation and evolution of disk galaxies show that outside the galactic
disk the mean metallicity of gas is $Z=0.2\pm0.1~Z_{\odot}$.  These
values of metallicities are also consistent with observational results
for the outskirts of groups \cite{Rasmussen2009} and clusters of
galaxies (e.g., \cite{Tamura2004}).  Thus it is highly unlikely that in
the CGM metallicity is as high as $Z_{\odot}$.  As such, 
$Z=Z_{\odot}$ sets a lower limit on the pathlength and
an upper limit on the electron density. The $1\sigma$ limits on the
pathlength and density are $L>41.6$ kpc and $n_e<2.6\times10^{-4}$
cm$^{-3}$.  For $Z=0.3Z_{\odot}$, which is far more likely, the
path-length becomes as large as $L=(239\pm100)$ kpc.

We can also estimate the total baryonic mass traced by the \ovii
absorbers, assuming a homogeneous spherically symmetric system:

\begin{equation}
M_{total}=(2.3\pm2.1)\times 10^{11}(\frac{f_c}{0.72}) (\frac{8.51 \times
10^{-4}}{(A_O/A_H)})^{3} 
(\frac{0.3~Z_{\odot}}{Z})^{3} (\frac{0.5}{f_{O VII}})^5 ~\msun
\end{equation}

for $L=239$ kpc and $f_{c}$ is the covering factor, $\sim72$\% for our
entire sample.

  For the $1\sigma$ lower limit on the path-length ($L>139$ kpc), 
the mass is $M_{total}>6.1 \times 10^{10} \msun$.  For the Solar 
abundance used by
\cite{Tumlinson11} $M_{total}= 1.2 \times 10^{10} \msun$, compared to
$2\times 10^9 \msun$ found by \cite{Tumlinson11} in \ovi systems. Thus
we find that the \ovii/\oviii systems probe the reservoir of gas in the
CGM extending to over 100 kpc. The mass probed by this warm-hot gas is
larger that that in any other phase of the CGM and is comparable to the
entire baryonic mass of the Galactic disk $\sim 6\times 10^{10} \msun$
\cite{SL06}. The baryonic fraction $f_{b}$ of this warm-hot gas varies
from $0.09-0.23$ depending on the estimates of the virial mass of the
Milky Way, from $10^{12}\msun$ to $2.5 \times 10^{12}\msun$ 
(\cite{Anderson2010} and references therein), bracketing the 
Universal value of $f_{b}=0.17$.

The oxygen mass in the CGM as traced by \ovii/\oviii is:

\begin{equation}
M_{oxygen}=(6.8\pm5.9) \times 10^{8} (\frac{0.5}{f_{O VII}})\msun
\end{equation}

and with the effective oxygen yi
eld $\sim0.01$ for a Salpeter IMF, the
stellar mass needed to produce this amount of oxygen is $M_{\star}=(6.8\pm5.9)
\times 10^{10}\msun$, which is of the order of the disk+bulge stars.

\subsection{Comparison with models}

\cite{Ntormousi2010} calculate all sky \ovii column density
distributions for halos of three Milky Way-like disk galaxies, resulting
from cosmological high-resolution N-body/gasdynamical simulations.
These simulations predict the mean logN(\ovii) ranging from
$13.46$--$14.55$, which is significantly lower than the average column
density we find.  Our observations, however, are consistent with those
of other such studies \cite{Fang2006, Williams2005, Williams2007,
Bregman2007}. We note that in the "pure hot halo" models of
\cite{Ntormousi2010}, the Local Group contribution is not included; this
might be a part of the reason for the discrepancy (see section $4.2$).

Our path-length and density measurements are in excellent agreement with
the recent estimates of the distribution of hot $(T\sim10^{6}~K)$
circum-galactic gas, based on high resolution cosmological
hydrodynamical simulations of \cite{Feldmann2012}.  These simulations
predict the presence of ionized circum-galactic gas with a density of
$\sim10^{-4}~cm^{-4}$ out to 100 kpc and beyond.  Hydrodynamic
simulations of CGM at $z = 0$ by \cite{Stinson11} found that their
simulated galaxies contain metal enriched warm-hot gas extending to
approximately the virial radius ( $250~kpc$). While their simulations do
not track the \ovii or \oviii producing gas, it is of interest to note
that the mass in the CGM is several times higher than the stellar mass
of their simulated galaxies, similar to what we find.  They argue that
the missing baryons exist in the CGM.


\subsection{Assumptions and Biases}

We have combined our measurements of the \ovii and \oviii column
densities with the published values of emission measure (EM) to derive
physical conditions in the absorbing/emitting plasma, viz.  temperature,
density, path-length and mass. The constraint on the temperature is
robust; it depends simply on the \ovii to \oviii column density ratio,
under the assumption of collisional ionization equilibrium. For our
measured density of a few times $10^{-4} cm^{-3}$, which is several
thousand times larger than the mean density of the Universe
($n=1.88\times 10^{-7} cm^{-3}$), photoionization becomes unimportant
and collisional ionization is a reasonable assumption.

We have assumed that the absorbing/emitting plasma is of uniform
density; this need not be the case. The distribution of warm-hot gas
around several galaxies follows a beta-model profile in which the
density is high in the center and falls off with radius (see
\cite{Mathur2008} and references therein); gas in clusters and groups of
galaxies also follows similar profile. In any case, if the gas is not of
uniform density, the EW would be weighted by the denser gas and for the
same observed column density the average density would be smaller and
the inferred path-length would be larger extending into the LG medium.
We cannot rule out the possibility that the warm-hot gas we trace is
from the LG medium instead. We note, however, that in the simulations of
\cite{Feldmann2012}, the density is roughly constant above the
Galactic disk out to about 100 kpc.

As noted above, our sample is made of sight-lines in which both \ovii and
\oviii K$\alpha$ absorption lines at $z=0$ are securely detected. Does
this mean that our sample is biased toward high column density systems?
If so, the path-length and the mass of the warm-hot plasma could be
severely over-estimated. In figure 3 we have plotted the EW distribution
of \ovii line in all 29 sources in the parent sample. The sub-sample of 8
sources is marked by the red shaded region; it is clear that this sub-sample
is {\it not} biased toward high EW systems. The high EW systems are not
in the sub-sample of this paper because they do not have clean detections
of \oviii lines, in large part because of the contamination from intrinsic
absorption. It is also worth noting that the upper limits on the
EWs in targets with non-detections of \ovii K$\alpha$ line is higher than the
average of the parent sample in all but one cases. All together, we
argue that the average EW of our sub-sample is a fair representation of
the large sample and that our sample is {\em not} biased toward high
column density systems.

\section{Conclusions}

For reasonable values of parameters and with reasonable
assumptions, the \chandra observations of \ovii and \oviii absorption
lines at $z=0$ imply that there is a huge reservoir of ionized gas
around the Milky Way. It may be in the halo of of the Milky Way or in
the surrounding Local Group. Either way, its mass appears to be very large.

\acknowledgements

Support for this work was provided by the National Aeronautics and Space
Administration through Chandra Award Number TM9-0010X issued by the
Chandra X-ray Observatory Center, which is operated by the Smithsonian
Astrophysical Observatory for and on behalf of the National Aeronautics
Space Administration under contract NAS8-03060.

\begin{figure}
\epsscale{.80}
\plotone{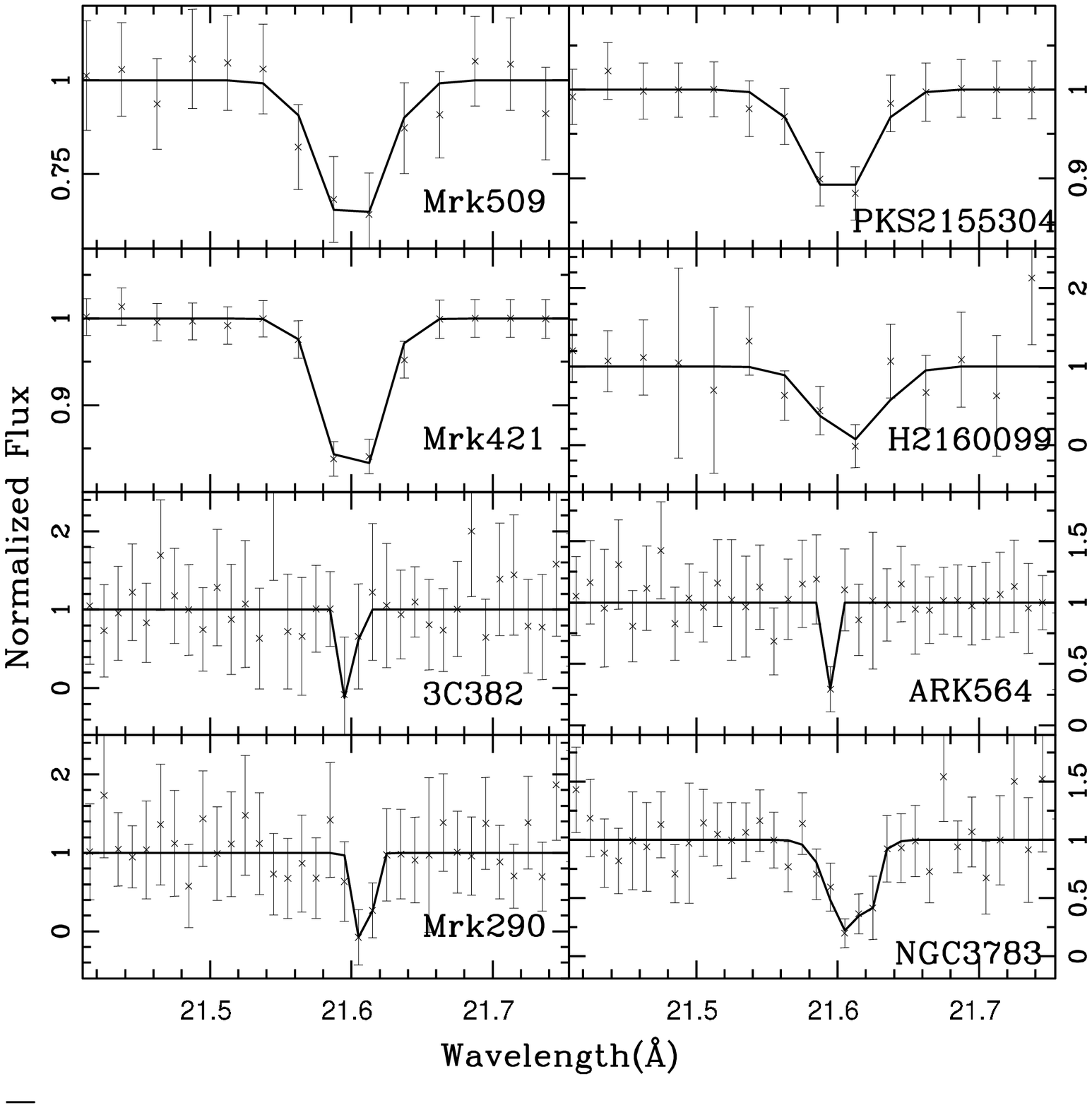}
\caption{The normalized flux at the location of the \ovii line at $21.602~$\AA.
}
\end{figure}

\clearpage

\begin{figure}
\epsscale{.80}
\plotone{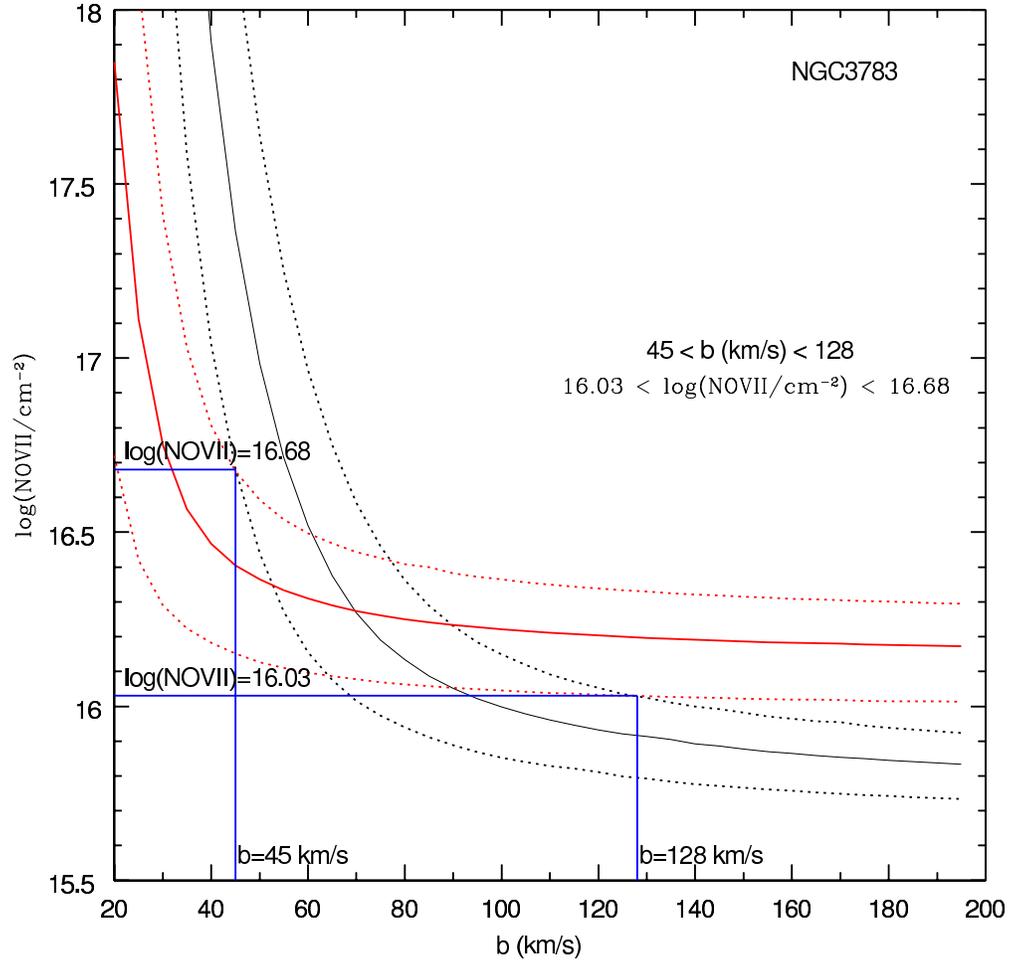}
\caption{Contours of allowed column densities N(\ovii) and Doppler parameters $b~$  for the \ovii$_{\rm K\alpha}$ (black) and \ovii$_{\rm K\beta}$ (red).
}
\end{figure}

\clearpage

\begin{figure}
\epsscale{.80}
\plotone{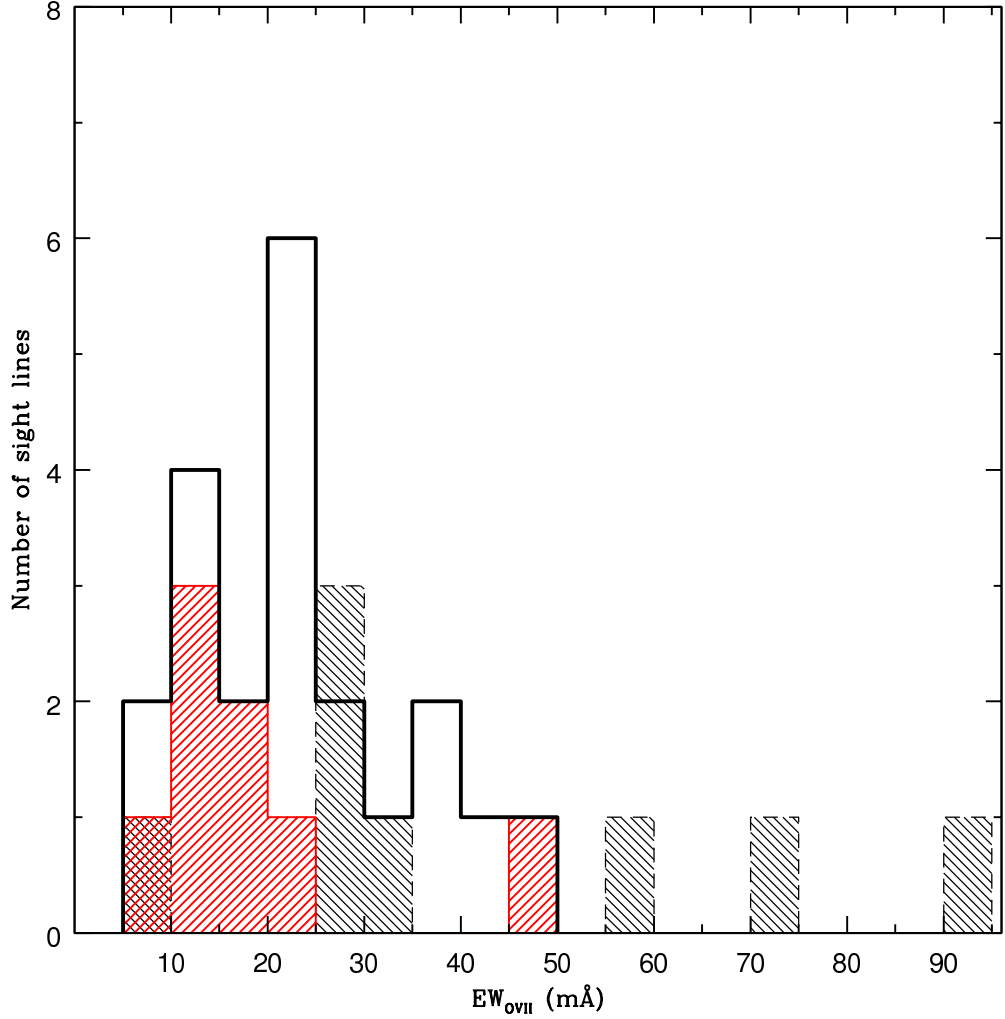}
\caption{Distribution of \ovii$_{\rm K\alpha}$ line EW for the 
parent sample. The solid black line corresponds 
to distribution for the 21 observations in which 
\ovii$_{\rm K\alpha}$ lines are clearly visible. 
The black shaded region marked the 
$1\sigma$ upper limits on the \ovii$_{\rm K\alpha}$ EW for 8 observations
with no (or less than $1\sigma$ significance) detection. The red shaded
region corresponds to sub-sample selected for this study (see text).
}
\end{figure}

\clearpage

\begin{table}[tp]\footnotesize
\caption{Summary of the targets used in this investigation.}
\begin{tabular}{ lcccccc }
\hline
Target & l & b & Redshift & Exposure\\
 & $deg$ & $(deg)$ & $z$ & $(ks)$\\
\hline
Mrk290 & 91.48 & 47.95 & 0.0304 & 250\\
PKS2155-304 & 17.73 &	-52.24 & 0.1160 & 530\\
Mrk421 & 179.83 & 65.03 & 0.0300 & 720 \\
Mrk509 & 35.97 & -29.86 & 0.0344 & 460 \\
3C382 &	61.30 & 17.44 & 0.0579 & 120 \\
Ark564 & 92.13 & -25.33 & 0.0247 & 250 \\
NGC3783 & 287.45 & 22.94 & 0.0097 & 905 \\
H2106-099 & 40.26 & -34.93 & 0.0265 & 100 \\
\hline
\end{tabular}
\end{table}

\clearpage

\begin{table}[tp]\footnotesize
\caption{The \ovii and \oviii absorption line measurement.}
\begin{tabular}{ lcccccc }
\hline
Target & EW (\ovii$_{\rm K\alpha}$) & EW (\ovii$_{\rm K\beta}$) & 
EW (\oviii$_{\rm K\alpha}$) & \ovii($\frac{EW ({\rm K\beta})}
{EW ({\rm K\alpha})}$) & \b & log(N\ovii)\\
 & $(m\AA)$ & $(m\AA)$ & $(m\AA)$ &  & $km/s$ & $cm^{-2}$\\
\hline
Mrk290 & $18.9\pm4.5$  & $5.1\pm3.7$ & $8.4\pm2.9$ & $0.27\pm0.21$  
& $>55$ & $16.14\pm0.32^{*}$\\
PKS2155-304 &    $11.6\pm1.6$  & $4.2\pm1.3$ & $6.7\pm1.4$ & 
$0.36\pm0.12$ & $35-94$ & $16.09\pm0.19$\\
Mrk421 &   $9.4\pm1.1$  & $4.6\pm0.7$ & $1.8\pm0.9$ 
& $0.49\pm0.09$   & $24-55$ & $16.22\pm0.23$\\
Mrk509 &  $23.9\pm5.0$  & $11.7\pm4.1$ & $10.3\pm4.3$ & 
$0.49\pm0.20$    & $70-200$ & $16.7\pm0.27$ \\
3C382 &	   $17.3\pm5.0$  & $7.8\pm3.0$ & $6.8\pm3.8$ & 
$0.45\pm0.22$   & $>40$ & $16.50\pm0.49^{*}$ \\
Ark564 &  $12.0\pm1.9$  & $<3.8$ & $9.5\pm4.1$ & 
$...$   & $>20$ & $15.82\pm0.20^{*}$\\
NGC3783 &  $14.4\pm2.5$  & $5.6\pm1.6$ & $4.5\pm2.9$ & 
$0.39\pm0.13$    &  $50-130$ & $16.30\pm0.25$ \\
H2106-099 &  $48.3\pm18.0$  & $<34.2$ & 
$28.8\pm13.8$ & $...$   & $>70$ & $16.23\pm0.16^{*}$\\
\hline
\end{tabular}
\end{table}
\normalsize{$^{*}$The lower limit on \ovii column densities are calculated 
using the curve-of-growth analysis.}\\

\end{document}